\documentclass[twocolumn,showpacs,amsmath,amssymb,aps,prl,superscriptaddress]{revtex4}
\usepackage{epsfig,psfrag,amsmath,amssymb,float}
\usepackage{color}
\input{epsf}

\newcommand{\bc}{\begin{center}}
\newcommand{\ec}{\end{center}}
\newcommand{\be}{\begin{equation}}
\newcommand{\ee}{\end{equation}}
\newcommand{\beqn}{\begin{eqnarray}}
\newcommand{\eeqn}{\end{eqnarray}}

\def\1.2{\frac{1}{2}}

\begin{document}
\title{Boundary effects in the critical scaling of entanglement
 entropy in 1D systems}
\author{Nicolas Laflorencie}
\affiliation{Department of Physics \& Astronomy, University of
British Columbia, Vancouver, B.C., Canada, V6T 1Z1}
\author{Erik S. S\o rensen}
\affiliation{Department of Physics and Astronomy, McMaster
University, Hamilton, ON, L8S 4M1 Canada}
\author{Ming-Shyang Chang}
\affiliation{Department of Physics \& Astronomy, University of
British Columbia, Vancouver, B.C., Canada, V6T 1Z1}
\author{Ian Affleck}
\affiliation{Department of Physics \& Astronomy, University of
British Columbia, Vancouver, B.C., Canada, V6T 1Z1}
\date{\today}
\begin{abstract}
We present exact diagonalization and density matrix renormalization group 
results for the entanglement entropy of critical spin-1/2 
XXZ chains. We find that 
open boundary conditions induce an alternating term in both 
the energy density and the entanglement entropy which are approximately  
proportional, decaying away from the boundary with a power-law.  The 
power varies with anisotropy along the XXZ critical line and is 
corrected by a logarithmic factor, which we calculate analytically, 
at the isotropic point. 
A heuristic resonating valence bond explanation is suggested.
\end{abstract}
\maketitle

{\it{Introduction ---}}
Quantum spin chains have recently acquired the status of paradigmatic
systems to investigate and understand the subtle interplay between quantum
entanglement and quantum criticality~\cite
{Osterloh02,Nielsen02,Vidal03,Cardy04,Refael04}. While 
much theoretical work in this area has focussed on periodic boundary 
conditions (PBC), experimental systems typically have open boundary 
conditions (OBC).  In this letter, we study a new effect of 
OBC on quantum entanglement in antiferromagnetic chains.

The  \textit{entanglement entropy} (EE), first introduced by
Bennett \textit{et al.}~\cite{Bennett96} in a quantum information context,
has recently received more attention from a condensed matter point of view
since it has been shown to exhibit a universal scaling for some one 
dimensional (1D)
quantum critical points~\cite{Vidal03,Cardy04,Wilczek94,Refael04}. 
Defined as the von Neumann entropy of
the reduced density matrix $\hat{\rho}(x)$ of a given subsystem of size $x$
embedded in a larger closed ring of size $L$, a universal scaling is
expected at a quantum critical point in the limit $x\gg 1$ and $L\to \infty $:
$
{\mathcal{S}}(x)=-\mathrm{Tr}{\hat{\rho}}\ln {\hat{\rho}}=\frac{c}{3}\ln
x+s_{1}$.
The prefactor $c$ is the central charge of the associated conformal field
theory (CFT)~\cite{Cardy04,Wilczek94} and $s_{1}$ is a non universal constant related
to the ultra-violet cut-off~\cite{noteXX}. The divergence
of ${\mathcal{S}}(x)$ with $x$ reflects an interesting property of the reduced
density matrix: ${\hat{\rho}}(x)$ has
a number of non negligible eigenvalues which also diverges with the
subsystem size, like $M(x)\sim {\rm{e}}^{{\mathcal{S}}}\sim
  x^{c/3}$. 

This critical scaling  has been verified numerically~\cite{Vidal03} 
by diagonalizing free-fermion type Hamiltonians
for the quantum Ising chain ($c=1/2$) or the XX chain ($c=1$). 
An important extension to critical
and non-critical systems with finite size, finite temperature and different
boundary conditions has been achieved by Calabrese and Cardy~\cite{Cardy04}.
Using CFT, they showed for instance that for critical systems of finite size
$L$ with OBC, the expression for the EE of a
subsystem of size $x$ including the open end becomes:
\begin{equation}
{\mathcal{S}}(x,L)=\frac{c}{6}\ln \left( \frac{2L}{\pi }\sin (\frac{\pi x}{L}%
)\right) +\ln g+s_{1}/2,  \label{eq:2}
\end{equation}
where $\ln g$ is boundary entropy introduced in \cite{IanBE}. 
(We have corrected factors of 2 in the last 2 terms in Eq. (\ref{eq:2})
 following \cite{Zhou}.) 

In the following, we investigate the EE for a large class of
critical XXZ quantum spin chains of length $L$ with OBC. In addition 
to confirming the behavior (1), at large $x$, we find an unexpected 
{\it{alternating term}} in the EE which decays away 
from the boundary with a power law. The boundary also introduces 
a slowly  decaying alternating term in the energy density.  We show 
numerically that these are proportional. 
  The
microscopic origin of the even-odd alternation of the EE will be
understood through a qualitative resonating valence bond (RVB) picture.

{\it{Critical chains with open ends ---}}
The Hamiltonian for XXZ quantum spin chains of length $L$ with OBC is
\begin{equation}
{\mathcal{H}}_{\Delta}=\sum_{x=1}^{L-1}\left[ \frac{1}{2}%
(S_{x}^{+}S_{x+1}^{-}+S_{x}^{-}S_{x+1}^{+})+\Delta
S_{x}^{z}S_{x+1}^{z}\right] .  \label{eq:XXZ}
\end{equation}
The critical regime is achieved for $|\Delta |\le 1$.
The energy density
$\left\langle h_{x}\right\rangle =\left\langle
(S_{x}^{+}S_{x+1}^{-}+S_{x}^{-}S_{x+1}^{+})/2+\Delta
S_{x}^{z}S_{x+1}^{z}\right\rangle $ is
 independent of $x$ in periodic chains. On the other hand, an open end breaks translational symmetry and
the energy density picks up a slowly decaying alternating term 
or ``dimerization'':
\begin{equation}
\langle h_{x}\rangle = E_U(x)+(-1)^xE_A(x), \label{eq:D1}
\end{equation}
where $E_A(x)$ 
becomes non zero near the boundary and decays slowly away from it.
 We can calculate $E_A(x)$ by Abelian
bosonization methods modified by OBC \cite{Marston}. 
The low energy effective Hamiltonian is simply that of a free massless 
relativistic boson, $\phi (x)$ with a  boundary condition, $\phi (0)=$ 
constant. 
$E_U(x)$ goes to a constant like $%
1/x^{2}$ (for all $|\Delta|\le 1$) 
but this is not of interest here. The alternating part 
of $h_x$ 
 is $(-1)^{x+1}\sin (\sqrt{4\pi K}\phi )$ where $K=%
\frac{\pi }{2(\pi -\cos ^{-1}\Delta )}$ is the Luttinger liquid parameter.
Using the standard conformal mapping we obtain in a finite
system of length $L$%
\begin{equation}
 E_A(x,L) \propto 
\left\langle \sin (%
\sqrt{4\pi K}\phi )\right\rangle \propto \frac{1}{\left[ \frac{L}{\pi }%
\sin \left( \frac{\pi x}{L}\right) \right] ^{K}}.  \label{Dimerization}
\end{equation}

At the Heisenberg [SU(2) invariant] point, $\Delta =1$, the 
low energy effective Hamiltonian contains a marginally irrelevant 
coupling constant, $\lambda$.
It leads to a contribution to the
  scaling dimension of O($\lambda$) for most operators \cite{Cardy86,Affleck}.  
The staggered energy density, $\sin \sqrt{2\pi}\phi$ has a scaling dimension \cite{Affleck}
\be \gamma = 1/2+\pi \sqrt{3}\lambda + O(\lambda^2).\ee 
This implies~\cite{Affleck}
that the bulk correlation function $G\equiv \langle h_ih_j\rangle$ decays as 
$1/[|i-j|(\ln|i-j|)^{3/2}]$.   With a boundary, one can derive 
a renormalization group equation for $E_A(x)$:
\be [\partial /\partial (\ln x ) + \beta (\lambda)\partial /\partial \lambda
 +\gamma (\lambda )]E_A(x)=0,\ee
which is the same as that obeyed by $G(x)$ except that the 
term $\gamma$ gets replaced by $2\gamma$. (Since we are 
interested in $E_A(x)$ for $x\gg 1$, the presence of the 
boundary does not affect the scaling dimension of
$\sin (\sqrt{2\pi}\phi)$. This 
can be seen from considering the operator product expansion 
of the marginal operator with $\sin (\sqrt{2\pi}\phi)$. 
At short distances, it is unaffected by the boundary.  On the 
other hand, corrections to 
scaling dimensions of {\it boundary} operators due 
to the marginal operator {\it are} affected non-trivially by the 
boundary condition.  See \cite{Qin}.)
This implies
$E_A(x) \propto 1/[\sqrt{|x|}(\ln|x|)^{3/4}]$. 
It is highly non-trivial to include both the log corrections 
and the finite size effects in $E_A(x,L)$. However, there 
{\it is} a simple result at $x=L/2$.  
Including the cubic term in the $\beta$-function for the 
marginal coupling constant, and other higher order corrections 
\cite{Barzykin},
this becomes:
\be E_A(\frac{L}{2},L)=a_0  
\frac{1+a_{2}/[\ln (L/a_{1})]^{2}
}{\sqrt{L}[\ln (L/a_{1})+(1/2)\ln \ln (L/a_{1})]^{\frac{%
3}{4}}}  ,  \label{logcorr}
\ee
where $a_0$, $a_1$ and $a_2$ are constants.
 This formula is fit 
to density matrix renormalization group (DMRG) data, (after extracting uniform and alternating part by 
fitting both locally to a polynomial using a 5-point formula)
in Fig. {\ref{fig:DimXXX}, obtaining excellent agreement.
In the same figure we also show $E_A(L/2)$ for the isotropic 
model with first and second neighbor interactions with $J_2/J_1=0.241167$. 
This model is at the critical point between gapless and gapped 
spontaneously dimerized phase and the marginal coupling constant, 
and hence the log corrections are expected to vanish here~\cite{Affleck}, as 
is seen in the figure.  
\begin{figure}[!ht]
\begin{center}
\includegraphics[width=\columnwidth]{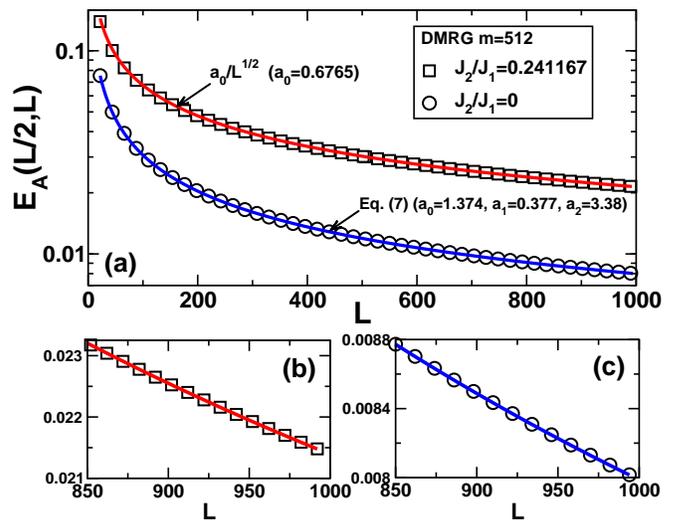}
\end{center}
\caption{$E_A(L/2,L)$ computed by DMRG (keeping $m=512$ states) at
  the SU(2) point up to $L=1000$ for the nearest neighbor model $J_2/J_1=0$ ($\circ$) 
  and at the critical point $J_2/J_1=0.241167$ ($\square$) (a). 
Full lines are fits, indicated on the main panel. Zooms on large $L$
  regions are showed in (b) and (c): (b) absence of log corrections
  for $J_2/J_1=0.241167$; (c) excellent agreement between Eq.~(7) and
  DMRG.} 
\label{fig:DimXXX}
\end{figure}

{\it{RVB picture ---}} This modulation in $\langle h_x\rangle$ can be
interpreted as an alternation of \textit{strong} and \textit{weak} bonds
along the chain. Indeed, the boundary spin at $x=1$ will have a strong
tendency to form a singlet pair with its only partner on the right hand
side. On the other hand the spin located at $x=2$ will be consequently less
entangled with its right partner at $x=3$ since it already shares a strong
entanglement with its left partner. 
\begin{figure}[th]
\begin{center}
\includegraphics[width=\columnwidth,clip]{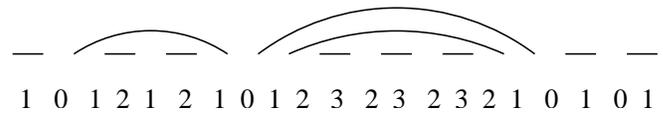}
\end{center}
\caption{A particular valence bond state.  The number of valence 
bonds crossing each link, $N(x)$, is given.}
\label{fig:RVB}
\end{figure}
This tendency towards dimer
formation over odd bonds induced by OBC can be naively depicted
through the simple valence bond solid state:
\begin{equation}
|\Phi \rangle =\frac{1}{2^{L/2}}\otimes_{x=1}^{L/2}[\mid \uparrow
_{2x-1}\downarrow _{2x}\rangle -\mid \downarrow _{2x-1}\uparrow _{2x}\rangle
].
\end{equation}
 (This is  the exact ground state of 
the Majumdar-Ghosh model with a second neighbor coupling obeying 
$J_2/J_1=1/2$ \cite{MG}.)
In such a state, the EE of a finite
subsystem alternates between 
$\ln 2$ if the interface between the subsystem and the rest of the chain cuts
a singlet bond and $0$ otherwise. More generally, we may write any 
singlet state as a linear combination of all valence bond states 
(with no bonds crossing each other). 
 It is trivial to calculate the EE 
for a simple state corresponding to any particular valence 
bond configuration, such as the one drawn in Fig. (\ref{fig:RVB}).
 It is given exactly by $S(x)=N(x)\ln 2$, 
where $N(x)$ is the number of valence bonds crossing the link $x$ between 
sites $x$ and $x+1$. 
Since one may think, in a renormalization 
group sense, of the ground state of 
the random-bond Heisenberg model as consisting of a single valence 
bond configuration~\cite{Fisher94}, this leads \cite{Refael04} to an interesting 
prediction for the EE. Unfortunately, the 
non-random critical  case, where the ground 
state involves an infinite sum over different valence bond configurations, 
including bonds of arbitrary length,  
 appears more difficult. Nevertheless, it seems plausible that the 
enhanced tendency towards valence bonds on odd links induced by a boundary 
will translate into an alternating term in the EE. 

{\it{Numerical results for the entropy---}}
It is well known that when the Ising exchange term $\Delta =0$, using a
Jordan-Wigner mapping, the XXZ Hamiltonian (\ref{eq:XXZ}) is equivalent to a
free fermion Hamiltonian
\begin{equation}
{\mathcal{H}}_{\mathrm{XX}}=\frac{J}{2}\sum_{x=1}^{L-1}\left[ \Psi
_{x}^{\dagger }\Psi _{x+1}+\Psi _{x+1}^{\dagger }\Psi _{x}\right],
\label{eq:FF}
\end{equation}
which can be solved in  momentum space. The computation of the
EE in this non-interacting case can be easily achieved
numerically for very large systems since it only requires 
diagonalization of an $%
L\times L$ matrix (see Ref.~\cite{Peschel} for some details). We first
report results from exact diagonalizations of the free fermion Hamiltonian (%
\ref{eq:FF}) with $L=2000$ sites. 
\begin{figure}[th]
\begin{center}
\includegraphics[width=\columnwidth]{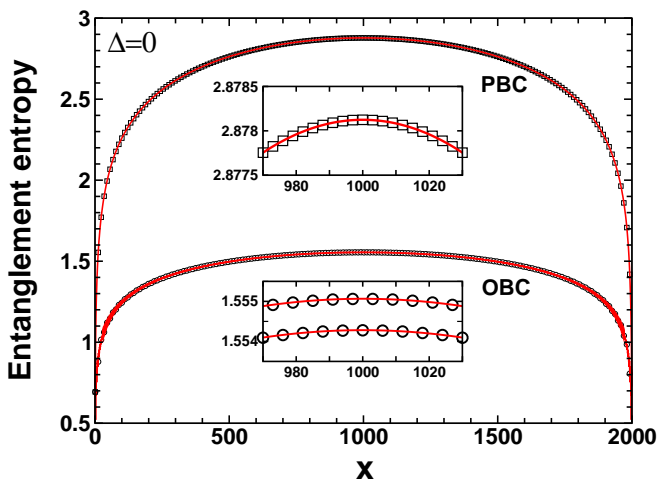}
\end{center}
\caption{Results from exact diagonalization of XX chains with $L=2000$
  spins $1/2$. EE ${\mathcal{S}}(x,2000)$, plotted vs the
  subsystem size $x$ in PBC (upper symbols) and OBC (lower
  symbols); the insets shows zooms around the chain center. The full
  lines are fits: Eq.~(\ref{PBC}) with $s_1=0.726067$ for the PBC case whereas in the OBC
  case, some uniform and staggered terms $\frac{0.055-(-1)^x0.25}{\frac{L}{\pi }\sin
\left( \frac{\pi x}{L}\right)}$ have been added to Eq.~(\ref{eq:2}).}

\label{fig:XX}
\end{figure}
In Fig.~\ref{fig:XX} we 
superimpose both PBC and OBC cases. In the PBC situation, as predicted in
Ref.~\cite{Cardy04}, the entropy is very well described by the following
expression
\begin{equation}
{\mathcal{S}}^{\mathrm{PBC}}(x,L)=\frac{c}{3}\ln \left( \frac{L}{\pi }\sin
\left( \frac{\pi x}{L}\right) \right) +s_{1},
\label{PBC}
\end{equation}
with $c=1$ and $s_{1}\simeq 0.726$~\cite{Jin04}. On the other hand, the open chain does
not obey the expression (\ref{eq:2}), but as expected from the RVB
picture presented above, an alternating term is found,
 as is visible in the lower inset of Fig.~\ref{fig:XX}. 
Indeed, the \textit{strong odd bond}
- \textit{weak even bond} picture agrees qualitatively  with an enhanced
(reduced) entropy for subsystems cutting an odd (even) bond.
We thus define the uniform and alternating parts of the EE
for large $x$:
\be \mathcal{S}(x,L)= \mathcal{S}_U(x,L)+(-1)^x\mathcal{S}_A(x,L),\ee
where both $\mathcal{S}_U(x,L)$ and $\mathcal{S}_A(x,L)$ are expected 
to be slowly varying functions for $x\gg 1$. 
 Interestingly, for the XX case, ${\mathcal{S}}_{A}(x,L)$ is found to decay slowly like a
power-law,
\begin{equation}
{\mathcal{S}}_{A}^{\rm{XX}}(x,L)\sim \frac{1}{\frac{L}{\pi }\sin (\frac{\pi
    x}{L})},
\end{equation}
which has the same exponent as the alternating part of the energy density 
$E_A$, as follows from Eq. (\ref{Dimerization})
 using 
the correct value $K=1$, for the $XX$ model. 
It turns out that both the alternating part of the EE
${\mathcal{S}}_{A}$ and ${E_A}$,
defined by
Eq.~(\ref{eq:D1}), are nearly proportional at large $x$ 
with a proportionality constant of 
$\pi /2$: $\mathcal{S}_{A}(x,L)=(-\pi /2){E_A}(x,L)+O(1/x^2)$.
${\mathcal{S}}_U(x,L)$ has also a boundary-induced correction 
to the result in Eq. (\ref{eq:2}), which is also $\propto 1/[L\sin (\pi x/L)]$.
This is included in the fit in Fig.~\ref{fig:XX}. 

\begin{figure}[ht!]
\begin{center}
\includegraphics[width=\columnwidth]{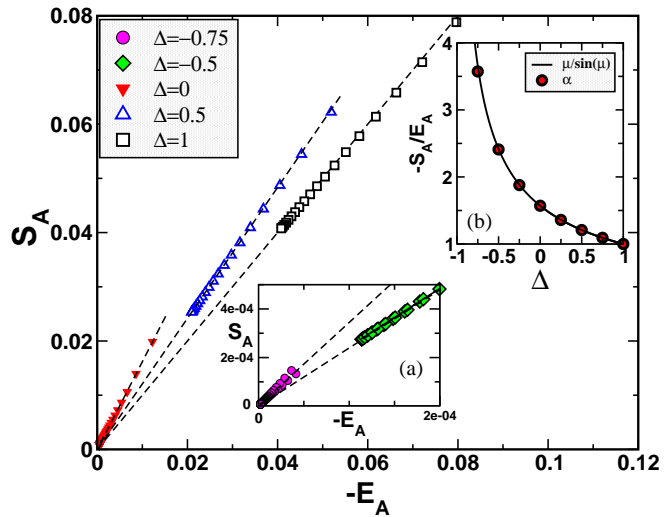}
\par
\end{center}
\caption{Linear behavior of the alternating part of the EE, ${\cal{S}}_A$ as 
a function of the alternating energy density, 
 $-E_A$, both computed using DMRG  on
critical open XXZ chains of size $200\le L\le 1000$ for a few values 
of the anisotropy $\Delta$. Data from ED at $\Delta=0$ are also shown for $L=2000$.
Dashed lines are linear fits of the form  ${\cal{S}}_A=-\alpha
E_A$ (see text).
The inset (a) is a zoom close to 0, showing data for $\Delta=-3/4$ and $-1/2$.
The inset (b) shows the prefactor $\alpha$ vs
$\Delta$ extracted from the numerical data (circles), for a larger 
set of values of $\Delta$,  which is compared with 
 $\pi /2v=\mu/\sin \mu$, $\mu =\cos^{-1}\Delta$.} 
\label{fig:Os-Dim}
\end{figure}
 Based on DMRG
data obtained on critical open chains of size $200\le L\le 1000$, 
we find this proportionality is still 
true even when $\Delta \neq 0$ where
the decaying behavior of $E_A$ is controlled by $K$. 
More precisely, plotting ${\cal{S}}_A$ as a function of $-E_A$ for
various values of the anisotropy $\Delta$ in
Fig.~\ref{fig:Os-Dim}, we find a linear relation  ${\cal{S}}_A=-\alpha
E_A$ with a prefactor perfectly described by $\alpha=\mu/\sin
\mu$, as shown in the inset of Fig.~\ref{fig:Os-Dim}, with $\mu =\cos
^{-1}\Delta$. We note that the spin-wave velocity for the XXZ model 
is given by $v=\pi (\sin \mu )/(2\mu )$ so that we may write this 
relation as 
\be \mathcal{S}_A=-(\pi a^2/2v)E_A,\label{SE}\ee
where we have introduced the lattice spacing, $a$, to make the EE a dimensionless quantity
($E_A$ has dimensions 
of energy per unit length.).
  We emphasize that Eq. (\ref{SE}) 
even holds for the Heisenberg model with $\alpha=1$ where we find both 
$E_A(L/2,L)$ and $\mathcal{S}_A(L/2,L)$ display the same logarithmic 
corrections, Eq.~(\ref{logcorr}), as shown in Fig.~\ref{XXX}. The sum
$E_A(L/2,L)+\mathcal{S}_A(L/2,L)$ is found to rapidly decay as a power-law, with a power $\simeq 2.59$ (see Fig.~\ref{XXX}).
Interestingly, for the $SU(2)$ invariant model with 
   $J_1=1$ and $J_2=0.241167$, again linearity is observed, but with a
   prefactor $\alpha\simeq 1.001689$ not related to the spin velocity,
   $v$, which we have determined to be $v\simeq 1.1699$~\cite{Note}.
   Hence,  Eq.~(\ref{SE}) does not hold in this case. In Fig.~\ref{XXX} we also
   show the sum $E_A(L/2,L)+\alpha\mathcal{S}_A(L/2,L)$, for this model, which decays rapidly 
with a power $\simeq 2.56$.
\begin{figure}[ht!]
\begin{center}
\includegraphics[width=\columnwidth,clip]{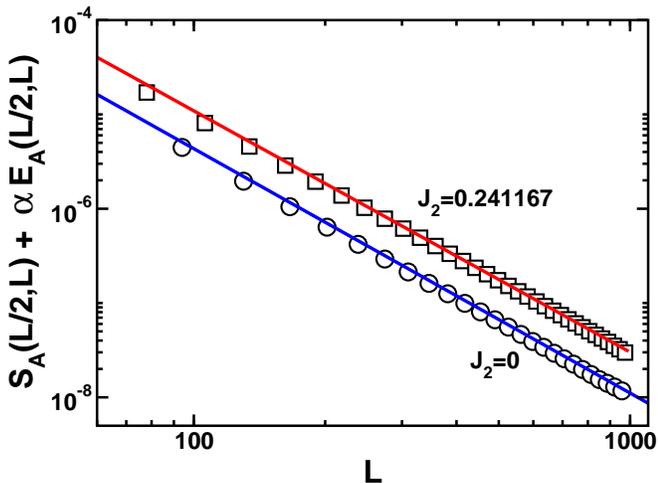}
\end{center}
\caption{
  Comparison
  between the alternating part of EE ${\cal{S}}_A$  and
 $E_A$ [Eq.~(\ref{eq:D1})] from DMRG with $m=512$ states, 
for Heisenberg models.
 Power-law decay of $\mathcal{S}_A(L/2,L)+\alpha E_A(L/2,L)$ 
drawn in a log-log plot, with $\alpha=1$ for the nearest neighbor chain ($J_2=0$) 
and $\alpha=1.00169$ at the critical second neighbor coupling $J_2=0.241167$. 
Lines are power-law fits: $\sim L^{-2.56}$ for $J_2=0.241167$ and $L^{-2.59}$ for $J_2=0$.}
\label{XXX}
\end{figure}

{\it{Conclusion ---}}
We emphasize that this alternating term 
in $\mathcal{S}(x,L)$ is universal and 
should {\it not} be regarded as a correction 
due to irrelevant operators.   First of all, it is not a ``correction'', 
since it is alternating. Secondly, it decays with the same power law
as $E_A(x,L)$ which is seen to be a property of the fixed point, 
not the irrelevant operators. (However, for the Heisenberg model, 
$\Delta =1$, the   log factor in $E_A(x,L)$ {\it is} due to the 
marginally irrelevant operator.)  The presence of a universal  
alternating term in $\mathcal{S}(x,L)$ 
is connected with the antiferromagnetic nature 
of the Hamiltonian (not appearing, for example, in the quantum Ising 
chain~\cite{Zhou})
and {\it does not} seem to follow from the general CFT 
treatment in \cite{Cardy04}. An analytic derivation of this 
phenomena remains an open problem. 

{\it{Acknowledgements ---}}
We are grateful to J. Cardy for interesting discussions. N.L. acknowledges 
I. Peschel for correspondence.
I.A. also wishes to thank S. Eggert for discussions about $E_A$. This 
research was supported by NSERC (all authors), the CIAR (I.A.) 
CFI (E.S.) and SHARCNET (E.S.).  
Numerical simulations have been performed on the WestGrid network and 
the SCHARCNET facility at McMaster University.


\begin{thebibliography}{99}
\bibitem{noteXX}  However, the constant contribution has been exactly
evaluated analytically for a critical XX chain: see~\cite{Jin04}.
\bibitem{Note} This velocity has been estimated from DMRG data using the finite size scaling 
of the ground state
 energy $E_0(L)=e_0L+e_1-\frac{\pi v}{24L}+O(L^{-2})$ on open 
chains up to $L=500$.
\end{thebibliography}
\end{document}